\documentclass{article}
\usepackage{dcolumn}
\usepackage{bm}
\usepackage{verbatim}       

\usepackage[dvips]{graphicx}
\usepackage{amsthm}
\usepackage{amssymb}
\usepackage{bm}
\usepackage{amstext}
\usepackage{amsmath}
\usepackage{amsfonts}
\newcommand{\p}{\partial}

\newcommand{\dd}{{\rm d}}
\newcommand{\bd}{\begin{definition}}                
\newcommand{\ed}{\end{definition}}                  
\newcommand{\bc}{\begin{corollary}}                 
\newcommand{\ec}{\end{corollary}}                   
\newcommand{\bl}{\begin{lemma}}                     
\newcommand{\el}{\end{lemma}}                       
\newcommand{\bp}{\begin{proposition}}            
\newcommand{\ep}{\end{proposition}}                
\newcommand{\bere}{\begin{remark}}                  
\newcommand{\ere}{\end{remark}}                     

\newcommand{\bt}{\begin{theorem}}
\newcommand{\et}{\end{theorem}}

\newcommand{\be}{\begin{equation}}
\newcommand{\ee}{\end{equation}}

\newcommand{\bit}{\begin{itemize}}
\newcommand{\eit}{\end{itemize}}
\newtheorem{theorem}{Theorem}[section]
\newtheorem{corollary}[theorem]{Corollary}
\newtheorem{lemma}[theorem]{Lemma}
\newtheorem{proposition}[theorem]{Proposition}
\theoremstyle{definition}
\newtheorem{definition}[theorem]{Definition}
\theoremstyle{remark}
\newtheorem{remark}[theorem]{Remark}

\begin{document}
%

\title{Global hyperbolicity is stable in the interval topology\thanks{This archive version includes a detailed discussion of Geroch's
original argument (see Sect. \ref{nhf}), which has been omitted in
the published version: J. Math. Phys.  {\bf 52} (2011) 112504. }}

\author{J.J. Benavides Navarro\thanks{Dipartimento di Matematica ``U. Dini'',
Universit\`a degli Studi di Firenze, Viale Morgagni 67/A, I-50134
Firenze, Italy. E-mail: navarro@math.unifi.it}  \ and E.
Minguzzi\thanks{ Dipartimento di Matematica Applicata ``G.
Sansone'', Universit\`a degli Studi di Firenze, Via S. Marta 3,
I-50139 Firenze, Italy. E-mail: ettore.minguzzi@unifi.it}}


\date{}
\maketitle

\noindent We prove that global hyperbolicity is stable in the
interval topology on the spacetime metrics. We also prove that every
globally hyperbolic spacetime admits a Cauchy hypersurface which
remains Cauchy under small perturbations of the spacetime metric.
Moreover, we prove that if the spacetime admits a complete timelike
Killing field, then the light cones can be widened preserving both
global hyperbolicity and the Killing property of the field.

\section{Introduction}
This work is devoted to the proof of the stability of the globally
hyperbolic property of spacetimes, where the word {\em stability}
refers to a particular topology on the space of spacetime metrics.
We stress that we are not going to consider the problem of  {\em
stability of a property under evolution}, indeed in this work we do
not impose any evolution equation, such as Einstein's equations, on
the spacetime. All the results of this work will  concern instead
the stability of a property of spacetime where the latter is
regarded as a whole.


Of course the interest of physicists for global hyperbolicity comes
from the fact that this property assures the predictability of the
spacetime evolution from initial data on a
 spacelike hypersurface. It is a property which allows one
to regard the spacetime geometrodynamics as deterministic, in
analogy with other field theories. Its  stability, in any considered
form, is therefore particularly welcomed as it shows that it could
be realized in nature. As mentioned, we shall consider only the
problem of stability with respect to {\em spacetime metric}
perturbations rather than {\em Cauchy data} perturbations.

In a sense this last problem, at least for global hyperbolicity,
makes less sense because, according to the Choquet-Bruhat--Geroch
theorem, for every vacuum initial data (given by a Riemannian space
and an extrinsic curvature tensor field $(S,h,K)$), there exists a
unique, up to isometric diffeomorphism, vacuum development $(M, g)$,
which is globally hyperbolic and inextendible in the class of
globally hyperbolic Lorentzian manifolds \cite{hawking73}.  Thus,
perturbing the initial conditions would still lead, after evolution,
to a globally hyperbolic spacetime which is still inextendible in
the class of globally hyperbolic spacetimes. In this evolutionary
sense global hyperbolicity is trivially stable. Of course, if one
pairs global hyperbolicity with the inextendibility property (not
restricting to the globally hyperbolic class) one obtains a
stability problem which is known as Strong Cosmic Censorship Problem
\cite{penrose69,penrose79,moncrief81}.



%

Coming to our framework, the topology on (conformal classes of)
spacetime metrics to which we shall refer to  is Geroch's interval
topology \cite{geroch70}. In this topology an open set is obtained
by giving, continuously and for each spacetime point, two light
cones one inside the other and by considering all the metrics whose
light cones respect those bounds (see next section). If a metric
stays in the open set then perturbing its coefficients in a
coordinate chart would keep the metric inside the open set, provided
the perturbation is sufficiently small, because the light cone on
the tangent space depends continuously on the metric coefficients
(indeed  the $C^0$ topology on metrics passes to the quotient of
conformal classes to Geroch's interval topology \cite{lerner73}).

Actually, Geroch's interval topology is one of the coarsest
topologies that can be given on the space of (conformal classes of)
metrics \cite{lerner73} and hence the stability in this topology is
particularly strong. There is only one other important  topology
that has been used in the literature and which is coarser than
Geroch's interval topology: the compact-open topology \cite[Sect.
6.4,7.6]{hawking73} \cite{hawking71}. In this topology a base open
set of the topology is obtained as above but the metric light cones
are bounded only inside a compact set of spacetime. Nevertheless, it
is obvious that a property such as global hyperbolicity cannot be
stable in this topology indeed, already for Minkowski $3+1$
spacetime, no matter how large a compact set $C$ is, and how much we
constraint the metric inside the compact set, it is possible to open
the cones sufficiently far from $C$ so as to produce closed causal
curves and hence spoil global hyperbolicity.

An argument of proof was produced by Geroch \cite{geroch70}, but, as
we shall explain in detail in Sect. \ref{nhf}, that argument does
not really work without introducing some non trivial amendments.
Perhaps for this reason, this result is not included in any textbook
devoted to mathematical relativity.

In the next section we give a  direct and strictly topological proof
of the stability of global hyperbolicity which, contrary to Geroch's
argument \cite{geroch70}, does not  use the concept of Cauchy
hypersurface or the topological splitting. As a corollary of this
result we obtain that every globally hyperbolic spacetime admits a
Cauchy hypersurface that remains Cauchy for small perturbations of
the spacetime metric (remark \ref{afh}).

We also prove  that if the spacetime admits a complete timelike
Killing field, then it is possible to enlarge the light cones while
keeping global hyperbolicity and stationarity in such a way that the
timelike Killing field is left unchanged.

We refer the reader to \cite{minguzzi06c} for most of the
conventions used in this work. In particular, we denote with $(M,g)$
a $C^r$ spacetime (connected, Hausdorff, time-oriented Lorentzian
and hence paracompact manifold), $r \in {3, \dots, \infty}$ of
arbitrary dimension $n \geq 2$ and signature $(-,+, \dots, +)$. On
$M \times M$ the usual product topology is defined. The subset
symbol $\subset$ is reflexive, thus $X \subset X$. We use the dot to
denote the boundary of a set, e.g. $\dot{X}$. With $J^+_{g}$ we
specify the causal relation referring to metric $g$. With $x<y$ we
mean that there is a future directed causal curve joining $x$ and
$y$ and we write $x\le y$ (also denoted $(x,y) \in J^{+}$) if $x<y$
or $x= y$. Given two metrics $g,g'$ we write $g<g'$ if, for every
$p\in M$, the causal cone of $g$ in $TM_p$ is contained in the
timelike cone of $g'$, and we write $g\le g'$ if for every $p\in M$,
the causal cone of $g$ in $TM_p$ is contained in the causal cone of
$g'$.

\section{The proof}

%

Let us recall that a spacetime is non-total imprisoning if no
inextendible causal curve is entirely contained in a compact set. A
globally hyperbolic spacetime $(M,g)$ is a causal spacetime such
that for every $p,q\in M$, $J^{+}(p)\cap J^{-}(q)$ is compact
\cite{bernal06b,minguzzi06c}. A spacetime is globally hyperbolic if
and only if it admits a Cauchy hypersurface, namely a (closed)
acausal set intersected by any inextendible  causal curve
\cite{hawking73,oneill83}. A useful alternative definition is
\cite{minguzzi08e}: a spacetime is globally hyperbolic if it is
non-total imprisoning and for every $p,q\in M$, $J^{+}(p)\cap
J^{-}(q)$ has compact closure.

A stably causal spacetime $(M,g)$ is a spacetime for which there is
a metric $g'>g$  such that $(M,g')$ is causal. Global hyperbolicity
implies stable causality which implies non-total imprisonment
\cite{hawking73,minguzzi06c}.
The intervals $(\underline{g},\overline{g})=\{g:
\underline{g}<{g}<\overline{g}\}$ form a base for Geroch's interval
topology on the space of conformal classes $Con(M)$ (i.e.
equivalence classes of metrics which are identical up to  a positive
conformal factor). This topology is equivalent to that induced from
 Whitney's fine $C^0$ topology on the space of metrics $Lor(M)$
(see \cite{lerner73}).

In general, a conformally invariant property $\mathcal{P}$ of
$(M,g)$ is said to be {\em stable in the interval topology} if there
is an interval $(\underline{g},\overline{g})\ni g$ of metrics that
share it. Since no mention is made of the differentiability of the
metrics contained the intervals this topology is quite coarse and
the stability with respect to it is a strong for of stability.

Some properties are inherited by spacetimes obtained by narrowing
the light cones, that is: if $(M,g)$ satisfies $\mathcal{P}$ and
$g''<g$ then $(M,g'')$ satisfies $\mathcal{P}$ (see \cite[Sect.
2]{minguzzi07}). One such such property is global hyperbolicity as
it is evident from the second definition given above. In this case
in order to prove the stability of the property we need only to show
that there is $g'>g$ which satisfies the property. In particular, we
shall look for a continuous $g'$. Of course, given a continuous
$g'>g$ which  does the job it is not difficult to find another
metric in the interval $(g,g')$ with the same degree of
differentiability of $g$. Ultimately, this is the reason why, in all
the considerations of stability, the differentiability of the
metrics will play no significative role.


The next result is well known, we include the proof for the sake of
completeness.

\begin{lemma} \label{one}
Let $(M,g)$ be a causal spacetime, then it  is globally hyperbolic
if and only if for every compact set $K$, $J^{+}(K)\cap J^{-}(K)$ is
compact.
\end{lemma}

\begin{proof}
One direction is trivial. For the other direction we have to prove
that $J^{+}(K)\cap J^{-}(K)$ is compact. Let us consider an
arbitrary sequence $p_n \in J^{+}(K)\cap J^{-}(K)$, then there are
two sequences $r_n,q_n \in K$ such that $r_n\le p_n\le q_n$. As a
first step let us show that $p_n$ is contained in a compact set,
which implies that $J^{+}(K)\cap J^{-}(K)$ has compact closure. If
not we could pass to a subsequence denoted in the same way such that
$p_n \to +\infty$ (i.e. escapes every compact set) while $r_n\to r$,
$q_n\to q$ for some $r,q\in K$. But then, taking $r'\ll r$ and
$q'\gg q$, we have since $I^{+}$ is open and $I^{+}\subset J^{+}$,
$p_n\in J^{+}(r')\cap J^{-}(q')$ which is compact, a contradiction.
We have therefore shown that there is $p\in M$ such that $p_n\to p$
and it remains to prove that $p \in J^{+}(K)\cap J^{-}(K)$. Again
passing to a subsequence if necessary we can assume $r_n \to r$ and
$q_n \to q$ for some $r,q\in K$. Since $J^{+}$ is closed (because
global hyperbolicity implies causal simplicity) we conclude $r \le p
\le q$ and hence the thesis.
\end{proof}

The next lemma is \cite[Lemma 3.2]{minguzzi07}.
\begin{lemma} \label{efs}
If $g<g'$ then $\overline{J^{+}_g}\subset J^{+}_{g'}$.
\end{lemma}

\begin{lemma} \label{nqz}
In a globally hyperbolic spacetime $(M,g)$, if $C$ is a compact set
and $\tilde{g}>g$ then $J^{+}(C)=
\bigcap_{g<g'<\tilde{g}}\overline{J^{+}_{g'}(C)}$ (and analogously
$J^{-}(C)= \bigcap_{g<g'<\tilde{g}}\overline{J^{-}_{g'}(C)}$).
\end{lemma}

\begin{proof}
Let us prove as a first step that
$\bigcap_{g<g'<\tilde{g}}{J^{+}_{g'}(C)}=
\bigcap_{g<g'<\tilde{g}}\overline{J^{+}_{g'}(C)}$. The inclusion
$\subset$ is obvious. For the other inclusion let us prove that for
every $\hat{g}$, $g<\hat{g}<\tilde{g}$,
$\bigcap_{g<g'<\tilde{g}}\overline{J^{+}_{g'}(C)} \subset
{J^{+}_{\hat{g}}(C)}$, the first step will follow from the
arbitrariness of $\hat{g}$. Indeed, let $q\in
\bigcap_{g<g'<\tilde{g}}\overline{J^{+}_{g'}(C)}$ and take
$\check{g}$, $g<\check{g}<\tilde{g}$, such that $\check{g}<\hat{g}$.
Since $q\in \overline{J^{+}_{\check{g}}(C)}$ there is a sequence
$q_n \in J^{+}_{\check{g}}(C)$ such that $q_n \to q$. Thus there are
$p_n\in C$ such that $(p_n,q_n)\in J^{+}_{\check{g}}$, and since $C$
is compact we can assume without loss of generality that $p_n \to
p\in C$. We conclude that $(p,q)\in \overline{J^{+}_{\check{g}}}$.
By lemma \ref{efs} we have $\overline{J^{+}_{\check{g}}}\subset
J^{+}_{\hat{g}}$ and hence $q \in J^{+}_{\hat{g}}(p)\subset
J^{+}_{\hat{g}}(C)$, or, due to the arbitrariness of $q$,
$\bigcap_{g<g'<\tilde{g}}\overline{J^{+}_{g'}(C)}\subset
J^{+}_{\hat{g}}(C)$, which completes the first step.

%

It remains to prove $J^{+}(C)=
\bigcap_{g<g'<\tilde{g}}{J^{+}_{g'}(C)}$. The inclusion $\subset$ is
trivial. For the other inclusion let $q\in
\bigcap_{g<g'<\tilde{g}}{J^{+}_{g'}(C)}$. This means that for every
$g'$, $g<g'<\tilde{g}$, $J^{-}_{g'}(q)\cap C\ne \emptyset$ and in
particular $\overline{J^{-}_{g'}(q)}\cap C\ne \emptyset$. However
(Lemma 3.3 of \cite{minguzzi07} or use \cite[Remark
3.8]{minguzzi07}), $J^{-}_S(q)= \bigcap_{g<g'<\tilde{g}}
\overline{J^{-}_{g'}(q)}$, thus $J^{-}_S(q)\cap
C=\bigcap_{g<g'<\tilde{g}} ( \overline{J^{-}_{g'}(q)}\cap C)\ne
\emptyset$ where $J^{+}_S:=\bigcap_{g'>g} J^{+}_{g'}$ is the Seifert
relation \cite{seifert71,minguzzi07}. Here the last equality follows
from the fact that we are taking the intersection of a family of
compact sets which satisfies the finite intersection property (the
finite intersection property follows from the fact that given a
finite family of metrics $>g$ there is one with the same property
and cones narrower  than any element of the family). In a globally
hyperbolic spacetime $J^+_S=J^+$ (this result follows from
\cite[Theorem 2.1]{hawking74}, or one can use the stronger results
\cite{minguzzi08b} that in a stably causal spacetime $J^{+}_S$ is
the smallest closed and transitive relation that contains $J^{+}$
and hence coincides with $J^{+}$ in globally hyperbolic spacetimes.
See also the appendix.). Thus $J^{-}_S(q)=J^{-}(q)$ from which $q\in
J^{+}(C)$ and the thesis follows.

\end{proof}

\begin{lemma} \label{hvr}
A globally hyperbolic spacetime $(M,g)$ admits a sequence of compact
sets $K_n$ with the properties (actually (i) follows from (ii))
\begin{itemize}
\item[(i)] $\bigcup_{n=0}^{\infty} K_n=M$,
\item[(ii)] If $C$ is a compact set then there is some $n\ge 0$ such that
$C\subset K_n$,
\item[(iii)] For $n\ge 0$, $J^{+}(K_n)\cap J^{-}(K_n) \subset
\textrm{Int}\, K_{n+1}$,
\item[(iv)] There is some metric $g_0>g$  such that $(M,g_0)$ is stably causal  and $\overline{J^{+}_{g_0}(K_0)}\cap \overline{ J^{-}_{g_0}(K_0)}
\subset \textrm{Int}\, K_{1}$
\end{itemize}
\end{lemma}

\begin{proof}
Since $M$ is a second-countable manifold it admits a complete
Riemannian metric $h$ \cite{nomizu61}. Chosen a point $w\in M$ the
closed balls $B(w, r)$ of $h$-radius $r\ge 0$ are compact by the
Hopf-Rinow theorem. Since the Riemannian distance is continuous it
reaches a maximum over a compact set, thus any compact set $C$ is
contained in some ball $B(w, r)$ for sufficiently large $r$. Let
$K_0=\{w\}$ and $r_0=1$ and define inductively
$K_{n+1}=J^{+}(B(w,r_n))\cap J^{-}(B(w,r_n))$ and $r_{n+1}$ to be
larger than $r_n+1$ and such that $ K_{n+1} \subset \textrm{Int}
B(w,r_{n+1})$. Clearly for $n\ge 1$, $J^{+}(K_n)\cap J^{-}(K_n)
\subset J^{+}(J^{+}(B(w,r_{n-1})))\cap
J^{-}(J^{-}(B(w,r_{n-1})))\subset K_n$ thus $J^{+}(K_n)\cap
J^{-}(K_n)=K_n$. For $n=0$ the last equality follows by causality
using $K_0=\{w\}$. Thus for $n\ge 0$, $J^{+}(K_n)\cap J^{-}(K_n)=K_n
\subset \textrm{Int} B(w,r_{n})\subset \textrm{Int} K_{n+1}$.
Furthermore since $r_{n+1}\ge r_n +1$ property (ii) is satisfied.
Finally, global hyperbolicity implies stable  causality and stable
causality is stable in the interval topology \cite[Lemma
2.2]{minguzzi07} thus there is $g_0>g$ such that $(M,g_0)$ is stably
causal; furthermore
$\overline{J^{\pm}_{g_0}(K_0)}=\overline{J^{\pm}_{g_0}(w)}={J^{\pm}_{g_0}(w)}$,
where we have used the closure of $J^{+}$ in a globally hyperbolic
spacetime. By causality $\overline{J^{+}_{g_0}(K_0)}\cap \overline{
J^{-}_{g_0}(K_0)}=\{w\} \subset \textrm{Int} B(w,1)\subset
\textrm{Int} K_1$.
\end{proof}

\begin{lemma} \label{ujh}
Given $n\ge 0$ assume there is some metric $g_n>g$ such that
$\overline{J^{+}_{g_n}(K_n)}\cap \overline{ J^{-}_{g_n}(K_n)}
\subset \textrm{Int}\, K_{n+1}$ then there is $g_{n+1}>g$ such that
$g_{n+1}\le g_n$, $g_{n+1}=g_n$ on $K_n$, and
$\overline{J^{+}_{g_{n+1}}(K_{n+1})}\cap \overline{
J^{-}_{g_{n+1}}(K_{n+1})} \subset \textrm{Int} \,K_{n+2}$.
\end{lemma}

\begin{proof}
By the properties satisfied by the set $K_n$ and by lemma \ref{nqz}
we have
\begin{align*}
\emptyset&=J^{+}(K_{n+1})\cap J^{-}(K_{n+1})\cap \dot{K}_{n+2} \\
&= \bigcap_{g<g'< g_n} \left( \overline{J^{+}_{g'}(K_{n+1})}\cap
\overline{ J^{-}_{g'}(K_{n+1})} \cap \dot{K}_{n+2}\right).
\end{align*}
If the compact sets on the right-hand side were non-empty they would
satisfy the finite intersection property, and thus also the
intersection would be non-empty. The contradiction proves that there
is a metric $g_{n+1}'$, $g<g_{n+1}'< g_n$, such that
\[
\overline{J^{+}_{g_{n+1}'}(K_{n+1})}\cap \overline{
J^{-}_{g_{n+1}'}(K_{n+1})} \cap \dot{K}_{n+2}=\emptyset.
\]
This equation implies that no $g_{n+1}'$-causal curve from $K_{n+1}$
can reach $\dot{K}_{n+2}$  and end at $K_{n+1}$ otherwise there
would be some point in ${J^{+}_{g_{n+1}'}(K_{n+1})}\cap {
J^{-}_{g_{n+1}'}(K_{n+1})} \cap \dot{K}_{n+2}$. We conclude that
${J^{+}_{g_{n+1}'}(K_{n+1})}\cap { J^{-}_{g_{n+1}'}(K_{n+1})}
\subset \textrm{Int} K_{n+2}$. Redefining $g_{n+1}'$ to be in the
interval $(g,g_{n+1}')$ and using lemma \ref{efs} one  obtains
$\overline{J^{+}_{g_{n+1}'}(K_{n+1})}\cap \overline{
J^{-}_{g_{n+1}'}(K_{n+1})} \subset \textrm{Int} K_{n+2}$.

We can now widen $g_{n+1}'$ on $\textrm{Int} \,K_{n+1}$ so as to
make it equal to $g_n$ on $K_n$  by defining $g_{n+1}=\chi
g_{n+1}'+(1-\chi) g_n$ where, $\chi: M \to [0,1]$, is a continuous
function such that $\chi=0$ on $K_n$ and $\chi=1$ on
$M\backslash\textrm{Int}\,K_{n+1}$. With this definition $g_{n+1}
\le g_n$ and we still have $\overline{J^{+}_{g_{n+1}}(K_{n+1})}\cap
\overline{J^{-}_{g_{n+1}}(K_{n+1})} \subset \textrm{Int} K_{n+2}$
because the metric has been widened only inside $\textrm{Int}\,
K_{n+1}$ so that
${J^{\pm}_{g_{n+1}}(K_{n+1})}={J^{\pm}_{g_{n+1}'}(K_{n+1})}$.

\end{proof}

\begin{theorem} \label{st}
Global hyperbolicity is stable in the interval topology.
\end{theorem}

\begin{proof}
We have to prove that on the globally hyperbolic spacetime $(M,g)$
there is some $g'>g$ such that $(M,g')$ is globally hyperbolic.

In the globally hyperbolic spacetime $(M,g)$ there is a sequence of
compact sets $K_n$ as in lemma \ref{hvr} and a sequence of metrics
$g_n>g$, such that $g_{n+1}\le g_n$ and
$\overline{J^{+}_{g_n}(K_n)}\cap \overline{ J^{-}_{g_n}(K_n)}
\subset \textrm{Int}\, K_{n+1}$. Indeed, condition (iv) in lemma
\ref{hvr} allows us to define inductively the sequence $g_n$ thanks
to lemma \ref{ujh}. We then define $g'(p)=g_n(p)$ if $p\in K_n$ so
that $g'>g$. The fact that $g_{n+1}=g_{n}$ on $K_n$ proves that $g'$
is continuous. Furthermore, since for every $i$, $g_{i+1}\le g_i$,
we have $g'\le g_i$ for every $i$. The fact that $g'\le g_0$ proves
that $(M,g')$ is stably causal (see lemma \ref{hvr} point (iv)) and
hence non-total imprisoning. In particular,
$\overline{J^{+}_{g'}(K_n)}\cap \overline{ J^{-}_{g'}(K_n)}
\subset\overline{J^{+}_{g_n}(K_n)}\cap \overline{ J^{-}_{g_n}(K_n)}
\subset \textrm{Int}\, K_{n+1}$, thus for every pair $p,q\in M$
there is some $K_n$ which contains $p$ and $q$ thus the set
$J^{+}_{g'}(p)\cap J_{g'}^{-}(q) \subset
\overline{J^{+}_{g'}(K_n)}\cap \overline{ J^{-}_{g'}(K_n)} \subset
K_{n+1}$ has compact closure.
We conclude that $(M,g')$ is globally hyperbolic.
\end{proof}

A time function $t:M\to \mathbb{R}$ is a continuous function such
that $x< y \Rightarrow t(x)< t(y)$. A Cauchy hypersurface $S$ is a
closed acausal set which is intersected by any inextendible causal
curve. Global hyperbolicity is equivalent to the existence of a
Cauchy hypersurface \cite{hawking73}.

We recall Geroch's topological splitting of globally hyperbolic
spacetimes \cite{geroch70} \cite[Prop. 6.6.8]{hawking73} which we
state in a slightly more detailed form (see remark \ref{sfb}).

\begin{theorem} \label{hdf}
Let $(M,g)$ be globally hyperbolic then there is a smooth manifold
$S$, a smooth projection $\pi:M\to S$, a time function $t:M\to
\mathbb{R}$, such that $\phi:=(t\times \pi): M \to \mathbb{R}\times
S$ is a homeomorphism with the property that each hypersurface
$S_a=\phi^{-1}(\{a\}\times S)$ is  $C^{1-}$ and Cauchy, and the
fibers $\pi^{-1}(s)$, $s\in S$, are ($C^1$) timelike curves.
Furthermore, given a smooth timelike vector field $v$, the fibers
$\pi^{-1}(s)$ can be chosen to be the integral lines of this field.

%
\end{theorem}

\begin{remark} \label{sfb}
The last statement,   the fact that the projection $\pi$ is actually
smooth in Geroch's splitting, and the fact that $S$ is a smooth
manifold are not so often noted.
It suffices to multiply the smooth timelike vector field $v$ by a
smooth spacetime function which makes it complete. The fact that the
quotient manifold is a smooth manifold and that $\pi$ is smooth is
then a standard result from manifold theory \cite[Theor.
9.16]{lee03}  as the integral lines of this field do not pass
arbitrarily close to themselves (because strong causality holds). We
shall need this fact only in the stationary case where it has been
already used, for instance in \cite{harris92}. Finally, we remark
that one could use also the fact that $\pi$ induces a homeomorphism
from $S_a$ to $S$ in order to define a smooth structure on $S_a$.
However, it would not be natural, as the inclusion of $S_a$ (endowed
with this smooth structure) in $M$ may be non-smooth.

%
\end{remark}

\begin{remark} \label{afh}
Thanks to theorem \ref{st} this result can be strengthened so that
the hypersurfaces $S_a$ are strictly acausal (namely there is $g'>g$
such that $S_a$ are acausal in $(M,g')$). In order to see this it
suffices to apply the previous theorem to $(M,g')$ where $g'>g$ is
such that $(M,g')$ is globally hyperbolic. Then, since every Cauchy
hypersurface for $(M,g')$ is a Cauchy hypersurface for $(M,g)$, one
gets easily the thesis. This observation is important as the
possibility of finding a strictly acausal Cauchy hypersurface in a
globally hyperbolic spacetime has been one of the main difficulties
behind the Cauchy hypersurface smoothability problem \cite{bernal03}
which essentially asks to prove that, not only the projection
$\pi:M\to S$, but also $t:M\to \mathbb{R}$ can be found smooth.
\end{remark}

%
%
%

\subsection{Stationary and static spacetimes}

Let $(M,g)$ be a spacetime admitting a timelike Killing vector field
$k$. We recall that the timelike Killing field $k$ is twist-free if,
defined the 1-form $\eta=g(\cdot,k)$, we have $\dd \eta\wedge
\eta=0$. This Frobenius condition is equivalent to the vanishing of
the vorticity tensor $w_{a b}=h^{c}_{\ a} h^{d}_{\ b}  u_{[c;d]}$
where $u=k/\sqrt{-g(k,k)}$ and $h^{a}_{\ b}=\delta^a_b+u^a u_b$ is
the projector on the vector space perpendicular to $u$ (the
equivalence follows from the fact that $w$ is orthogonal to $u$ thus
it vanishes if and only if $\varepsilon_{abcd} w^{ab} u^c$
vanishes).

The next result shows that in a globally hyperbolic spacetime it is
possible to widen the light cones preserving both global
hyperbolicity and stationarity (or staticity).

\begin{theorem}
Let $(M,g)$ be a globally hyperbolic spacetime admitting a complete
timelike Killing vector field $k$. There is a smooth function
$\alpha:M\to \mathbb{R}$, $L_k\alpha=0$, $\alpha>0$, such that
defined $g'=g-\alpha g(\cdot,k)\otimes g(\cdot,k)$, we have $g'>g$
and $(M,g')$ is globally hyperbolic. In particular $k$ is timelike
and Killing also for $(M,g')$. Finally, if $k$ is also twist-free
(hypersurface orthogonality, staticity) then $g'$ is such that the
property is preserved in $(M,g')$.
\end{theorem}

\begin{proof}
Let $\tilde{g}>g$ be such that $(M,\tilde{g})$ is globally
hyperbolic and let us consider Geroch's splitting of the spacetime
$(M,\tilde{g})$ as given in theorem \ref{hdf}, $\phi=(t\times
\pi):M\to \mathbb{R}\times S$, where $\pi:M\to S$ is the smooth
projection on a smooth quotient manifold such that the fibers are
the integral lines of $k$. The Killing field is indeed smooth, that
is, it reaches the largest differentiability degree allowed by the
differentiability properties of the spacetime manifold and the
metric, because for a Killing field: $k_{\beta;\alpha;\mu}=-R_{\mu
\nu \alpha \beta } k^\nu$.

Let $K_n$ be a sequence of compact sets on $S$ such that $K_n\subset
\textrm{Int} K_{n+1}$ and $S=\cup_n K_n$. Let $\varphi_r$ be the
one-parameter group of diffeomorphisms generated by $k$, and let
$A=\{x\in M: x=\varphi_r(s_0), \ s_0\in S_0, \vert r \vert < 1 \}$
where $S_0$ is the Cauchy hypersurface $t=0$ in Geroch's splitting.
In other words $A$ is the open set between $\varphi_{-1}(S_0)$ and
$\varphi_{1}(S_0)$. We remark that these latter hypersurfaces need
not be Cauchy for $\tilde{g}$. The sets $C_n=\overline{A}\cap
\pi^{-1}(K_n)$ are compact sets, indeed $C_n$ is a closed set
contained in $\phi^{-1}([-j,j]\times K_j)$ for sufficiently large
$j$, and $\phi=(t\times \pi)$ is a homeomorphism (more in detail:
for each $s_0\in \pi^{-1}(K_n)\cap S_0$ the segment of orbit
$\varphi_r(s_0), \vert r\vert\le 1$, is compact because of the
completeness of $k$ and thus the function $t$ reaches a maximum and
a minimum on it that depend continuously on $s_0$, then one uses
compactness of $\pi^{-1}(K_n)\cap S_0$). Let $g_{1/k}=g-\frac{1}{k}
g(\cdot,k)\otimes g(\cdot,k)$. By compactness and continuity for
each $n\ge 1$ there is some $k(n)$ such that $g_{1/k(n)}<\tilde{g}$
on $C_n$ and we can choose $k(n)$ to be an increasing function. Let
$\tilde{\alpha}: S\to (0,+\infty)$ be a smooth function such that
$\tilde{\alpha}<1/k(n)$ in $K_n\backslash \textrm{Int} K_{n-1}$.


Let us define $\alpha =\pi^{*}\tilde\alpha$, so that $\alpha :M\to
(0,+\infty)$ is smooth and satisfies $L_k\alpha=0$. The expression
$g' =g-\alpha g(\cdot,k)\otimes g(\cdot,k)$ satisfies $L_k g'=0$ on
$M$ and $g'<\tilde{g}$ on $\bar{A}$. In particular the spacetime $A$
with the metric induced from $g'$ is globally hyperbolic with Cauchy
hypersurface $S_0$.

Now we are going to prove that any past inextendible timelike curve
$\gamma$ in $(M,g')$ with a point $p$ in the region $t> 0$ must
actually intersect $t=0$ which, together with the dual statement
will provide a proof that $S_0$ is a Cauchy hypersurface for
$(M,g')$. Indeed, let $A_r=\varphi_r(A)$, so that $A_r$ is made by
the hypersurfaces  $\varphi_{r'}(S_0)$, for $r-1<r'<r+1$, then each
$(A_r,g'\vert_{A_r})$ is globally hyperbolic (simply because it is
isometric with $A$) with Cauchy hypersurface $\varphi_r(S_0)$ (again
by isometry). Let $\tau>0$ be such that $p=\varphi_\tau(s_0)$, with
$s_0\in S_0$, then we can find a finite increasing sequence $r_i\ge
0$, $i=0,1,\ldots m$, $r_0=0$, $r_m=\tau$ such that $r_{k+1}<r_k+1$.
Let us consider $A_{r_{m-1}}$. By construction $p$ belongs to
$A_{r_{m-1}}$ and stays in the chronological future of
$\varphi_{r_{m-1}}(S_0)$ thus $\gamma$ intersects
$\varphi_{r_{m-1}}(S_0)$ in a point which we denote $p_{m-1}$. Again
by construction $\varphi_{r_{m-1}}(S_0)\subset A_{r_{m-2}}$ thus
$\gamma$ intersects a point $p_{m-2}\in \varphi_{r_{m-2}}(S_0)$.
Continuing in this way we conclude, after a finite number of steps,
that $\gamma$ intersects $S_0$ which proves that  $(M,g')$ is
globally hyperbolic.

%
%
%
%
%
%
%
%
%
%
%


Finally, $g'(\cdot,k)=(1-\alpha g(k,k))g(\cdot,k)$ thus since
$(1-\alpha g(k,k))>0$ the kernel of the 1-form $g'(\cdot,k)$
coincides with the kernel of the 1-form $g(\cdot,k)$. The local
integrability of this kernel is the twist-free condition, hence the
thesis.
\end{proof}

\section{Criticism of Geroch's argument} \label{nhf}
In its influential paper \cite{geroch70} ``Domains of dependence''
after introducing the concept of domain of dependence and global
hyperbolicity, Geroch devoted a section to the proof of the
stability of global hyperbolicity. In this section we shortly review
Geroch's argument clarifying why and where it needs an amendment.

We start recalling  Geroch's construction of the time function in
his topological splitting theorem \cite[Sect. 3.7]{minguzzi06c}
\cite{hawking73}. Let $(M,g)$ be globally hyperbolic. The first step
is the introduction of a suitable finite volume measure on spacetime
(the measure $m$ of a complete Riemannian metric $h$ on $M$ would
work provided it is conformally rescaled to make $m(M)=1$). The
functions on $M$ defined by $t^{-}(p)=m(I^-(p))$ and
$t^{+}=-m(I^{+}(p))$ can be proved to be time functions. The former
goes to zero following any past inextendible curve, while the latter
goes to zero following any future inextendible curve. The time $t$
appearing in Geroch's topological splitting theorem is $t(p)=\ln
(-t^{-}(p)/t^{+}(p))$ and is surjective on $\mathbb{R}$. Geroch
proves that every surface $S_a=t^{-1}(a)$ is a Cauchy hypersurface
\cite[Prop. 6.6.8]{hawking73} and in order to prove the stability of
global hyperbolicity he splits the problem in two parts focusing
first in the spacetime region $D^{+}(S_0)$ and then in $D^{-}(S_0)$.
His idea is to show that it is possible to widen the cones
preserving the Cauchy property of $S_0$ on past inextendible curves
in $D^{+}(S_0)$. To do this he regards $D^{+}(S_0)$ has a countable
union of compact sets $K_n$, constructed as cups, whose explicit
form is $K_n=\overline{I^{-}({B}_n)\cap I^{+}(S_0)}$ where the
hypersurface $B_n$ is  defined by $B_n=\{p\in M: 1/n=m(I^{+}(p))\le
m(I^{-}(p))\}$. He then proves that for each $n$ there is some
$g_n>g$ such that $J^{+}_{g_n}(K_n)\cap J^{-}_{g_n}(K_n)=K_n$,
namely he proves that the compacts $K_n$ are stably causally convex.
Ultimately, it is the attempt at proving this particularly strong
result that causes most problems in Geroch's strategy (in our proof
we needed only the weaker result $\overline{J^{+}_{g_n}(K_n)}\cap
\overline{J^{-}_{g_n}(K_n)}\subset K_{n+1}$).
From this point the proof is not very detailed but it seems that
Geroch has in mind a gluing procedure similar to the one that can be
found in our proof.

\begin{figure}[ht]
\begin{center}
 \includegraphics[width=3cm]{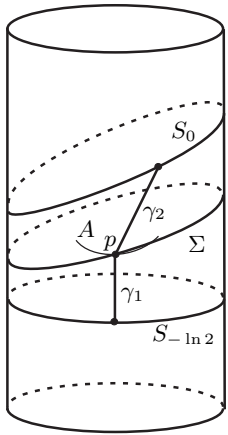}
\end{center}
\caption{The spacetime $\mathbb{R}\times S^1$ of metric $\dd
s^2=-\dd t^2+\dd \theta^2$. The timelike geodesic $\gamma_1$
maximizes the Lorentzian distance between $S_{-\ln 2}$ and $p$ while
the timelike geodesic $\gamma_2$ maximizes the Lorentzian distance
between $p$ and $S_{0}$. The set $A$ in Geroch's argument is
perpendicular to  $\gamma_1$ at $p$ and intersects $\Sigma$ in more
than one point.} \label{cil}
\end{figure}

The problem with his argument is that in order to prove the strong
result $J^{+}_{g_n}(K_n)\cap J^{-}_{g_n}(K_n)=K_n$ he has to assume
that $S_0$ and $B_n$ are strictly spacelike, namely locally strictly
acausal (for each $p \in S_0$ there is an open set $O\ni p$, such
that $S\cap O$ is acausal in $(O,g\vert_O)$). He thus follows a path
that inverts the logical development of our proofs in which our
remark \ref{afh} follows the main theorem \ref{st} on stability. The
reader may convince herself that the assumption that the Cauchy
hypersurfaces are strictly spacelike is particulary strong, in fact
as much as the original result on stability that one wishes to
prove. Indeed, Geroch  feels the need of giving  an argument which
would allow one to replace the hypersurfaces $S_0$ and $B_n$ with
strictly spacelike ones.

This argument is rather involved and in our opinion does not work.
We sketch how, according to Geroch, one would have to replace $S_0$.
He suggests to consider $S_{-\ln2}$ and $S_0$ and to replace $S_0$
with the hypersurface which is equidistant from  those according to
the Lorentzian distance (which in a globally hyperbolic spacetime is
continuous \cite{beem96,minguzzi08e}) $\Sigma=\{p\in D^{+}(S_{-\ln
2})\cap D^{-}(S_0): d(S_{-\ln 2},p)=d(p,S_0)\}$. His proof that
$\Sigma$ is strictly spacelike goes as follows. He takes $p\in
\Sigma\backslash\{S_0\cup S_{-\ln 2}\}$ and considers the maximal
(proper length) timelike geodesic $\gamma$ connecting $S_{-\ln 2}$
to $p$, i.e. $l(\gamma)=d(S_{-\ln 2},p)$. Then he takes a normal
neighborhood $O\ni p$ and a point $q\in \gamma\cap O$, so that $q\le
p$ and $d(q,p)=\epsilon>0$. The set $A=\{r:d(q,p)=\epsilon\}$
contains $p$ and by the reverse triangle inequality for every $r\in
A$, $d(S_{-\ln2},r)\ge l(\gamma)$. From this fact Geroch claims that
it follows that $\Sigma$ is strictly spacelike.
It seems that Geroch is assuming that $A$ intersects $\Sigma$ only
at $p$ while this is not the case because $\Sigma$ is not the locus
of points at distance $l(\gamma)$ from $S_{-\ln 2}$. As $p$ runs
over $\Sigma$ the length of the maximal geodesic varies. Of course,
if $A$ were entirely on the same side of $\Sigma$ then it would be
possible to open the cone at $p$, but this is not the case as figure
\ref{cil} shows. Furthermore, in order to prove the strict
spacelikeness of the hypersurface  one would have to prove that the
cones can be opened in a whole neighborhood of $p$, which could also
be troublesome as $\gamma$ might not vary continuously with $p$.

Geroch's way of replacing the hypersurfaces could be amended by
using the smoothability results contained in
\cite{bernal03,bernal04}. One would have to make this replacement
for all the Cauchy hypersurfaces appearing in Geroch's proof showing
that this step does not really affect the  argument. Of course the
proof would be no more self contained and would lose simplicity.
Our proof is instead more topological and does not even use the
notion of time function not to mention the topological splitting of
globally hyperbolic spacetimes.

\section{Conclusions}

We have given a topological proof of the stability of global
hyperbolicity in Geroch's interval topology. This result implies
that every spacetime admits a Cauchy hypersurface which remains
Cauchy for small perturbations of the spacetime metric. Moreover, we
have proved that the cones can be enlarged preserving not only
global hyperbolicity but also the presence of a complete timelike
Killing field, being it twist-free or not.


In our opinion the stability of global hyperbolicity  might prove to
be important for the investigation of the stability of geodesic
singularities and singularity theorems, \cite{lerner73} while its
role for predictability issues connected with the cosmic censorship
problem\cite{penrose79} seems less clear.

%
%

At the mathematical level it seems worthwhile to investigate the
analogous problem of the stability of causal simplicity. This
problem is harder because the property is not preserved by narrowing
the light cones, although, examples seem to suggest that causal
simplicity is indeed stable in the interval topology.


\section*{Acknowledgments}
We thank A. Fathi and A. Siconolfi for some useful discussions on
the issue of the stability of global hyperbolicity. Their
suggestions have allowed us to simplify the proof. This
work\footnote{(Note not included in the published version.) This
work was first submitted to Class. Quantum Grav. on 12th November
2010 and then rejected on 8th July 2011 (to be then rapidly accepted
by JMP). Apparently, the referee was very conservative and did not
judge positively the possibility of proving the stability of global
hyperbolicity without passing through smoothing arguments. We argued
that in fact the right approach should be the opposite, which we
advocate in remark \ref{afh}, namely to prove topologically that
global hyperbolicity is stable and then to prove, as a last step,
that smooth time functions exist. This strategy has been
successfully followed in the recent paper by A. Fathi and A.
Siconolfi: {\em On smooth time functions}, to appear in Math. Proc.
Camb. Phil. Soc., see e.g. Eq. (32) of that work.} has been
partially supported by GNFM of INDAM under project Giovani
Ricercatori 2009 ``Stable and generic properties in relativity''.

\section*{Appendix}
The Seifert relation is defined by $J^{+}_S=\bigcap_{g'>g}
J^{+}_{g'}$. In lemma \ref{nqz} we have used the equivalence
$J^{+}_S=J^{+}$ which holds in globally hyperbolic spacetimes. We
are going to give a proof of this fact which is similar to
\cite[Theorem 2.1]{hawking74}, but does not use the geodesic
triangulation.


We recall that in a globally hyperbolic spacetime the Lorentzian
distance $d:M\times M\to [0,+\infty]$ is finite, continuous and
maximized by a connecting geodesic \cite{beem96}.

\begin{theorem} \label{juj}
In a globally hyperbolic spacetime $J^{+}_S=J^{+}$.
\end{theorem}

\begin{proof}
Assume that: ``if $x \ll y$ then there is $g'>g$ such that
$J^{+}_{g'}(y) \subset I^{+}(x)$''. From this assumption it follows
$J^{+}_S(y)\subset I^{+}(x)\subset J^{+}(x)$. Taking the limit $x\to
y$ and using $\overline{J^{+}}=J^{+}$, $J^{+}_S(y)\subset J^{+}(y)$
the other inclusion being obvious.

It remains to prove $x \ll y \Rightarrow$ there is $g'>g$ such that
$J^{+}_{g'}(y) \subset I^{+}(x)$.




Let us define for $0<k<1$, $S_{k}=\{w: d(x,w)=k d(x,y)\}$. We have
that $I^{+}(S_k)=\{w: d(x,w)>k d(x,y)\}$, and $S_k=\p I^{+}(S_k)$,
in particular $S_k$ is an achronal boundary and $J^{+}(y)\subset
I^{+}(S_k)$, $S_k\subset I^{+}(x)$. It is  trivial to establish that
$S_k$ is actually acausal. We are going to find a strictly spacelike
hypersurface $\tilde{S}\subset I^{+}(S_{1/3})\cap I^{-}(S_{2/3})$
(see figure \ref{prhaw}). Using the results of \cite{bernal03} it
would be trivial since it is easy to establish that $S_{1/3}$ and
$S_{2/3}$ are Cauchy hypersurfaces for the spacetime $I^{+}(x)$, and
there is always a smooth spacelike hypersurface between two Cauchy
hypersurfaces, one in the future of the other \cite[Prop.
14]{bernal03}. We shall however give a self contained proof.

For every point $q\in S_{2/3}$ let $p(q)$ be the intersection
between a (make any choice) maximal geodesic $\sigma$ connecting $x$
to $q$ and $S_{1/3}$. By the reverse triangle inequality this
segment from $x$ to $p$ is also maximal and since $p\in S_{1/3}$, it
has length $d(x,y)/3$, thus the remaining part of the segment,
connecting $p$ to $q$ has length $d(x,y)/3$. For $\epsilon$,
$0<\epsilon<d(x,y)/3$ sufficiently small we can find a point
$r(q)\in \sigma$ at a distance $\epsilon$ from $q$ so that $r$ is
included in a relatively compact convex neighborhood $C_q$ of $q$.

Moreover, keeping $C_q$, there must be an even smaller $\epsilon$
such that ($r$ is at distance $\epsilon$ from $q$) $J^{+}(r)\cap
J^{-}(S_{2/3})\subset C_q$. Indeed if for every $\epsilon>0$ there
is some $s(\epsilon)\in J^{+}(r)\cap J^{-}(S_{2/3})\cap M\backslash
C_q$ then  connecting $r(\epsilon)$ to $s(\epsilon)$ with a maximal
geodesic, using the closure of $J^{+}$ in a globally hyperbolic
spacetime and taking the limit $\epsilon\to 0$ (which implies
$r(q)\to q$) we could find a point $\tilde{s}\in \dot{C}_q\cap
J^{-}(S_{2/3})$ in the causal future of $q$ which is a contradiction
with the acausality of $S_{2/3}$.


\begin{figure}[ht]
\begin{center}
 \includegraphics[width=9cm]{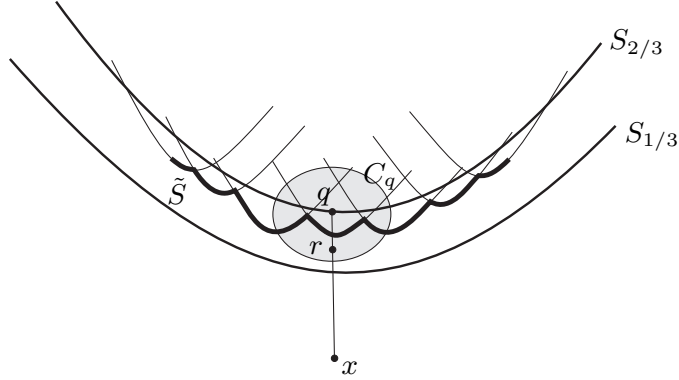}
\end{center}
\caption{The construction of the strictly spacelike hypersurface in
the proof of theorem \ref{juj}.} \label{prhaw}
\end{figure}

By the global hyperbolicity of $(M,g)$ as $S_{2/3}\cap
\overline{C_q}$ is compact $J^{+}(r) \cap J^{-}(S_{2/3})\subset
J^{+}(r)\cap J^{-}(S_{2/3}\cap \overline{C_q}$ and hence $J^{+}(r)
\cap J^{-}(S_{2/3})$ is compact.



Thus for sufficiently small $\epsilon$ the set $S_{2/3}\cap
I^{+}(r)$ is non-empty, as it contains $q$, and open relatively
compact in the topology of $S_{2/3}\cap C_q$ and furthermore
$J^{+}(r) \cap J^{-}(S_{2/3})$ is a compact set contained in $C_q$.
In particular $A(q)=\{w: d(r(q),w)= \epsilon/2 \} \cap
I^{-}(S_{2/3})\subset C_q$ is differentiable with spacelike tangent
by a property of the distance in a convex set (Gauss lemma
\cite[Lemma 4.5.2]{hawking73}).
 The open sets $O(q)=\{w: d(r(q),w)>d(r(q),q)/2 \}\cap S_{2/3}$
cover $S_{2/3}$. Since $S$ is $\sigma$-compact we can pass to a
locally finite countable covering $O(q_i)$.

(*) The set $\tilde{S}=\p I^{+}(\bigcup_i A(q_i))$ is a gluing of
the hypersurfaces $A(q_i)$. In the convex set $C_{q_i}$ the cones
can be opened keeping $A(q_i)$ spacelike and since locally only a
finite number of $C_q$ intersect, the same is true for $\tilde{S}$.
(It must be noted that by gluing the sets $A(q_i)$ instead of
summing the associated distance functions  from $r_i$ we don't have
to deal with some subtleties related to the fact that the squared
Lorentzian distance is not $C^1$ where it vanishes.)

(**) Alternatively, instead of using route (*) we can define  the
function $h_i(z)=[d(r_i,z)/d(r_i,q_i)]^4$ if $z\in J^{-}(S_{2/3})$
(which implies $z\in C_{q_i}$ or $h_i(z)=0$) and $h_i(z)=+\infty$ if
$z\in I^{+}(S_{2/3})$. The function $h_i$ is a $C^1$  time function
wherever it is  finite and furthermore it has a timelike gradient
wherever it is different from zero (it is sometimes said that $d^2$
is $C^1$ but this is false at the points where the Lorentzian
distance vanishes). In order to check that $h_{i}$ is $C^1$ it
suffices to use normal coordinates at $r$ and \cite[Lemma
5.9]{oneill83} according to which $h_{i}=[(x^0)^2-(x^1)^2\cdots-
(x^{n-1})^2]^2$. Defined $t=\sum_i h_i$ the function $t$ is a $C^1$
has (past directed) timelike gradient where it is different from
zero, and by construction (see the definition of $O(q)$) it takes on
$S_{2/3}$ a value greater than $1/2^4$. Thus at
$\tilde{S}=t^{-1}(1/2^5)$ the function $t$ is $C^1$ with a timelike
gradient.


For the last step, we find $g'>g$ such that $\tilde{S}$ is locally
$g'$-acausal. No $g'$-causal curve issued from  $y$ can cross
$\tilde{S}$ thus $J^{+}_{g'}(y)\subset I^{+}(S_{1/2}) \subset
I^{+}(x)$.
\end{proof}




\end{document}